\documentclass[livrev]{article}

\begin{document}
\bibliographystyle{livrev97}
\newcommand{\keywords}[1]{} 
\newcommand{\comment}[1]{}

\title{Numerical Approaches to Spacetime Singularities}

\author{Beverly K. Berger \\
Physics Department, Oakland University,
Rochester, MI 48309 USA \\ 
http://www.oakland.edu/\~{}berger \\
and \\
Physics Division, National Science Foundation \\
4201 Wilson Blvd., Arlington, VA 22230 USA \\
bberger@nsf.gov}

\date{}
\maketitle 

                      Abstract:
This Living Review updates a previous version \cite{berger98d} which its itself
an update of a review article
\cite{berger97a}. Numerical exploration of the properties of
singularities could, in principle, yield detailed understanding of their nature
in physically realistic cases. Examples of numerical investigations into the
formation of naked singularities, critical behavior in collapse, passage
through the Cauchy horizon, chaos of the Mixmaster singularity, and
singularities in spatially inhomogeneous cosmologies are discussed.
\keywords{numerical relativity, singularities}

\tableofcontents
\section{Introduction}
The singularity theorems \cite{wald84,hawking67,hawking73,hawking70} state that
Einstein's equations will not evolve generic, regular initial data arbitrarily
far into  the future or the past. An obstruction
such as infinite curvature or the termination of geodesics will always arise to
stop the evolution somewhere. The simplest, physically relevant solutions
representing for example a homogeneous, isotropic universe
(Friedmann-Robertson-Walker (FRW))
or a spherically symmetric black hole (Schwarz\-schild) contain space-like
infinite curvature singularities.  Although, in principle, the presence of a
singularity
could lead to unpredictable measurements for a physically realistic
observer, this does not happen for these two
solutions. The surface of last scattering of the cosmic microwave background in
the cosmological case and the event horizon in the black hole (BH) case
effectively hide the singularity from present day, external observers. The
extent to which this ``hidden'' singularity is generic and the types of
singularities that appear in generic spacetimes remain major open questions in
general relativity. The questions arise quickly since other exact solutions
to Einstein's equations have singularities which are quite different
from those described above. For example, the charged BH (Reissner-Nordstrom
solution) has a time-like singularity. It also contains a Cauchy horizon (CH)
marking the boundary of predictability of space-like
initial data supplied outside the BH. A test observer can pass through the CH to
another region of the extended spacetime. More general cosmologies can exhibit
singularity behavior different from that in FRW. The Big Bang in FRW is
classified as an asymptotically velocity term dominated (AVTD) singularity
\cite{eardley72,isenberg90} since any spatial curvature term in the Hamiltonian
constraint becomes negligible compared to the square of the expansion rate as
the singularity is approached. However, some anisotropic, homogeneous models
exhibit Mixmaster dynamics (MD) \cite{belinskii71b,misner69} and are not
AVTD---the influence of the spatial scalar curvature can never be neglected.
For more rigorous discussions of the classification and properties of the types
of singularities see \cite{ellis77,tipler80}.

Once the simplest, exactly solvable models are left behind, understanding of
the singularity becomes more difficult. There has been significant analytic
progress 
\cite{wald97,moncrief97,rendall97a,andersson00}. However, until recently
such methods have yielded either detailed knowledge of unrealistic, simplified
(usually by symmetries) spacetimes or powerful, general results that do not
contain details. To overcome these limitations, one might consider numerical
methods to evolve realistic spacetimes to the point where the properties of the
singularity may be identified. Of course, most of the effort in numerical
relativity applied to BH collisions has addressed the avoidance of
singularities \cite{finn97}. One wishes to
keep the computational grid in the observable region outside the horizon. Much
less computational effort has focused on the nature of the singularity itself.
Numerical calculations, even more than analytic ones, require finite values for
all quantities. Ideally then, one must describe the singularity by the
asymptotic non-singular
approach to it. A numerical method which can follow the evolution into
this asymptotic regime will then yield information about the singularity. Since
the numerical study must begin with a particular set of initial data, the
results can
never have the force of mathematical proof. One may hope, however, that such
studies will provide an understanding of the ``phenomenology'' of singularities
that will eventually guide and motivate rigorous results. Some examples of the
interplay between analytic and numerical results and methods will be given here.

In the following, we shall consider examples of numerical study of
singularities
both for asymptotically flat (AF) spacetimes and for cosmological models. These
examples have been chosen to illustrate primarily numerical studies whose focus
is the nature of the singularity itself. In the AF context, we shall consider
two questions:

The first is whether or not naked singularities exist for realistic matter
sources. 
One approach has been to explore highly non-spherical collapse looking
for spindle or pancake singularities. If the
formation of an event horizon requires a limit on the aspect ratio of the
matter \cite{thorne74}, such configurations may yield a naked singularity.
Analytic results suggest that one must go beyond the failure to observe an
apparent horizon to conclude that a naked singularity has formed
\cite{wald97}.
Another approach is to probe the limits between initial configurations which
lead to black holes and those which yield no singularity at all (i.e.~flat
spacetime plus radiation) to explore the singularity as the BH mass goes to
zero. This quest led naturally to the discovery of critical
behavior in the collapse of a scalar field \cite{choptuik93}. In the initial
study, the critical (Choptuik) solution is a zero mass naked singularity
(visible from null infinity). It is a counterexample to the cosmic censorship
conjecture
\cite{hamade96b}. However, it is a non-generic one since 
fine-tuning of the initial data is required to produce this critical solution.
In a possibly related study, Christodoulou has shown
\cite{christodoulou97} that for the spherically symmetric Einstein-scalar field
equations, there always exists a perturbation that will convert a solution with
a naked singularity (but of a different class from Choptuik's) to one with a
black hole.  Reviews of critical phenomena in gravitational collapse can
be found in \cite{bizon96,gundlach96a,gundlach98,gundlach99}. 

The second question which is now
beginning to yield to numerical attack involves the stability of the Cauchy
horizon in charged or rotating black holes. It has been conjectured
\cite{wald84,chandrasekhar82} that a real
observer, as opposed to a test mass, cannot pass through the CH since realistic
perturbed spacetimes will convert the CH to a strong spacelike singularity
\cite{tipler80}. Numerical studies \cite{brady95b,droz96,burko97a} show that a
weak, null singularity forms first as had been predicted \cite{poisson89,ori91}.

In cosmology, we shall consider both the behavior of the Mixmaster model
and the issue of whether or not its properties are applicable to generic
cosmological singularities.  Although numerical evolution of the Mixmaster
equations has a long history, developments in the past decade were motivated by
inconsistencies between
the known sensitivity to initial conditions and standard measures of the chaos
usually associated with such behavior
\cite{moser73,rugh90a,rugh90b,berger94,francisco88,burd90,hobill91,pullin91}.
A coordinate invariant characterization of Mixmaster chaos has
been formulated \cite{cornish97b} which, while criticized in
its details \cite{motter01}, has essentially resolved the question. In
addition, a new extremely fast and accurate algorithm for Mixmaster simulations
has been developed \cite{berger96c}. 

Belinskii, Khalatnikov, and Lifshitz (BKL)
long ago claimed
\cite{belinskii69a,belinskii69b,belinskii71a,belinskii71b,belinskii82} that it
is possible to formulate the generic cosmological solution to Einstein's
equations near the singularity as a Mixmaster universe at every spatial point.
While others have questioned the validity of this claim \cite{barrow79},
numerical evidence has been obtained for oscillatory behavior in the approach
to the singularity of spatially inhomogeneous cosmologies
\cite{weaver98,berger98a,berger98c,berger01}. We shall discuss results from a
numerical program to address this issue \cite{berger93,berger98c,berger01a}. The
key claim by BKL is that sufficiently close to the singularity, each spatial
point evolves as a separate universe---most generally of the Mixmaster type.
For this to be correct, the dynamical influence of spatial derivatives
(embodying communication between spatial points) must be overwhelmed by the time
dependence of the local dynamics. In the past few years, numerical simulations
of collapsing, spatially inhomogeneous cosmological spacetimes have provided
strong support for the BKL picture
\cite{berger93,berger97b,berger97e,weaver98,berger98a,berger98c,berger01}. In
addition, the Method of Consistent Potentials (MCP) \cite{grubisic93,berger98c}
has been developed to explain how the BKL asymptotic state arises during
collapse. New asymptotic methods have been used to prove that open sets exist
with BKL's local behavior (although these are AVTD rather than of the Mixmaster
type) \cite{isenberg98,kichenassamy98,andersson00}.  Recently, van Elst, Uggla,
and Wainwright developed a dynamical systems approach to $G_2$ cosmologies
(i.e.~those with 2 spatial symmetries) \cite{vanelst01}.

\section{Singularities in AF spacetimes}
While I have divided this topic into three subsections, there is considerable
overlap. The primary questions can be formulated as the cosmic censorship
conjecture. The weak cosmic censorship conjecture
\cite{penrose69} requires a singularity formed from regular, asymptotically
flat initial data to be hidden from an external observer by an event horizon.
An excellent review of the meaning and status of weak cosmic censorship has
been given by Wald \cite{wald97}. Counter examples have been known for a long
time but tend to be dismissed as unrealistic in some way. The strong form of
the cosmic censorship conjecture \cite{penrose79} forbids timelike
singularities, even within black holes.

\subsection{Naked singularities and the hoop conjecture}
\subsubsection{Overview}
Perhaps, the first numerical approach to study the cosmic censorship
conjecture consisted of attempts to create naked singularities. Many of these
studies were motivated by Thorne's ``hoop conjecture''
\cite{thorne74} that collapse will yield a black hole only if a mass $M$ is
compressed to a region with circumference $C
\le 4 \pi M$ in all directions. (As is discussed by Wald \cite{wald97}, one
runs into difficulties in any attempt to formulate the conjecture precisely.
For example, how does one define $C$ and $M$, especially if the initial data are
not at least axially symmetric? Schoen and Yau defined the
size of an arbitrarily shaped mass distribution in \cite{schoen83}. A
non-rigorous prescription was used in a numerical study by Chiba
\cite{chiba99}.) If the hoop conjecture is true, naked singularities may form if
collapse can yield $C \ge 4\pi M$ in some direction. The existence of a naked
singularity is inferred from the absence of an apparent horizon (AH) which can
be identified locally. Although a definitive
identification of a naked singularity requires the event horizon (EH) to be
proven to be absent, to identify an EH requires knowledge of the entire
spacetime. While one finds an AH within an EH \cite{israel86a,israel86b}, it is
possible to construct a spacetime slicing which has no AH even though an EH
is present
\cite{wald91}. Methods to find an EH in a numerically determined spacetime have
only recently become available and have not been applied to this issue
\cite{libson96,masso98}. A local prescription, applicable numerically, to
identify an ``isolated horizon'' is under development by Ashtekar et al (see
for example \cite{ashtekar00}).

\begin{figure}[bth]
\begin{center}
\caption{Heuristic illustration of the hoop conjecture. }
\end{center}
\end{figure}

\subsubsection{Naked Spindle Singularities?}
In the best known attempt to produce
naked singularities, Shapiro and Teukolsky (ST) \cite{shapiro91} considered
collapse of prolate spheroids of collisionless gas. (Nakamura and
Sa\-to~\cite{nakamura82} had previously studied the collapse of
non-rotating deformed stars with an initial large reduction of internal
energy and apparently found spindle or pancake singularities
in extreme cases.)
ST solved the general relativistic Vlasov equation for
the particles along with Einstein's equations for the gravitational field.
They then searched each spatial slice for trapped surfaces. If no trapped
surfaces were found, they concluded that there was no AH in that slice. The
curvature invariant $I =R_{\mu \nu \rho \sigma}R^{\mu \nu \rho
\sigma}$ was also computed. They found that an AH (and presumably a BH) formed
if $C \le 4\pi M < 1$ everywhere but no AH (and presumably a naked singularity)
in the opposite case. In the latter case, the evolution (not surprisingly)
could not proceed past the moment of formation of the singularity.
In a subsequent study, ST \cite{shapiro92} also showed that a
small amount of rotation (counter rotating particles with no
net angular momentum) does not prevent the formation of a naked
spindle singularity. 
However, Wald and Iyer \cite{wald91} have shown that the Schwarzschild solution
has a time slicing whose evolution approaches arbitrarily close to the
singularity with no AH in any slice (but, of course, with an EH in the
spacetime). This may mean that there is a chance that the increasing
prolateness found by ST in effect changes the slicing to one with no apparent
horizon just at the point required by the hoop conjecture. While, on the face
of it, this seems unlikely, Tod gives an example where a trapped surface does
not form on a chosen constant time slice---but rather different portions form at
different times. He argues that a numerical simulation might be
forced by the singularity to end before the formation of the trapped surface is
complete. Such a trapped surface would not be found by the simulations
\cite{tod92}. In response to such a possibility, Shapiro and Teukolsky
considered equilibrium sequences of prolate relativistic star clusters
\cite{shapiro93}. The idea is to counter the possibility that an EH might
form after the time when the simulation must stop. If an equilibrium
configuration is non-singular, it cannot contain an EH since singularity
theorems say that an EH implies a singularity. However, a sequence of
non-singular equilibria with rising $I$ ever closer to the spindle singularity
would lend support to the existence of a naked spindle singularity since one
can approach the singular state without formation of an EH. They constructed
this sequence and found that the singular end points were very similar to
their dynamical spindle singularity. Wald believes, however, that it is likely
that the ST slicing is such that their singularities are not naked---a trapped
surface is present but has not yet appeared in their time slices \cite{wald97}. 

Another numerical study of the hoop conjecture was made by Chiba
et al \cite{chiba94}. Rather than a dynamical collapse model, they searched
for AH's in analytic initial data for discs, annuli, and rings. Previous
studies of this type were done by Nakamura et al \cite{nakamura88} with oblate
and prolate spheroids and by Wojtkiewicz \cite{wojtkewicz90} with
axisymmetric singular lines and rings.
The summary of their results is that an AH forms if $C \le 4\pi M \le
1.26$. (Analytic results due to Barrab\`{e}s et al
\cite{barrabes91,barrabes92} and Tod \cite{tod92} give similar quantitative
results with different initial data classes and (possibly) an
alternative definition of $C$.) 

There is strong analytical evidence against the development of spindle
singularities. It has been shown by Chru\'sciel and Moncrief
that strong cosmic censorship holds in AF electrovac solutions which admit a
$G_2$ symmetric Cauchy surface \cite{berger95a}. The evolutions of these highly
nonlinear equations are in fact non-singular. 

\subsubsection{Recent results}
Garfinkle and Duncan \cite{garfinkle01a} report preliminary results on
the collapse of prolate configurations of Brill waves \cite{brill59}.
They use their axisymmetric code to explore the conjecture of Abrahams et
al \cite{abrahams92} that prolate configurations with no
AH but large $I$ in the initial slice will evolve to form naked
singularities. Garfinkle and Duncan find that the configurations become
less prolate as they evolve suggesting that black holes (rather than naked
singularities) will form eventually from this type of initial data. Similar
results have also been found by Hobill and Webster \cite{hobill01}.

Pelath et al \cite{pelath98} set out to generalize previous results
\cite{wald91,tod92} that formation of a singularity in a slice with no
AH did not indicate the absence of an EH. They looked specifically at
trapped surfaces in two models of collapsing null dust, including the
model considered by Barrab\`{e}s et al \cite{barrabes91}. They indeed
find a natural spacetime slicing in which the singularity forms at the
poles before the outermost marginally trapped surface (OMTS) (which defines the
AH) forms at the equator. Nonetheless, they also find that whether or not
an OMTS forms in a slice closely (or at least more closely than one would
expect if there were no relevance to the hoop conjecture) follows the
predictions of the hoop conjecture.

\begin{figure}[bth]
\begin{center}
\caption{This figure is based on Fig.~1 of \protect \cite{pelath98}. The
vertical axis is time. The blue curve shows the singularity and the red
curve the outermost marginally trapped surface. Note that the singularity forms
at the poles (indicated by the blue arrow) before the outermost marginally
trapped surface forms at the equator (indicated by the red arrow). }
\end{center}
\end{figure}

\subsubsection{Going further}
Motivated by ST's results \cite{shapiro91}, Echeverria \cite{echeverria93}
numerically
studied the properties of the naked singularity that is known to form in the
collapse of an infinite, cylindrical dust shell \cite{thorne74}. While
the asymptotic state can be found analytically, the approach to it must
be followed numerically. The analytic asymptotic solution can be matched
to the numerical one (which cannot be followed all the way to the collapse)
to show that the singularity is strong (an observer experiences infinite
stretching parallel to the symmetry axis and squeezing perpendicular to
the symmetry axis). A burst of gravitational radiation
emitted just prior to the formation of the singularity
stretches and squeezes in opposite directions to the singularity. This result for
dust conflicts with rigorously nonsingular solutions for the electrovac case
\cite{berger95a}. One wonders then if dust collapse gives any information about
singularities of the gravitational field.

One useful result from dust collapse has been the study of gravitational
waves which might be associated with the formation of a naked
singularity. Such a program has been carried out by Harada, Iguchi, and
Nakao \cite{iguchi98,iguchi99,nakao00,harada00a,harada00b,iguchi01}. 

Nakamura et al (NSN) \cite{nakamura93} conjectured that
even if naked spindle singularities could exist, they would either
disappear or become black holes. This demise of the naked singularity
would be caused by the back reaction of the gravitational waves emitted
by it. While NSN proposed a numerical test of their conjecture, they
believed it to be beyond the current generation of computer technology.

Chiba \cite{chiba99} extended previous results \cite{chiba94} to search
for AH's in spacetimes without axisymmetry but with a discrete symmetry.
The discrete symmetry is used to identify an analog of a symmetry axis to
allow a prescription for an analog of the circumference. Given this
construction, it is possible to formulate the hoop conjecture in this
case and to explore its validity in numerically constructed momentarily
static spacetimes. Explicit application was made to multiple black holes
distributed along a ring. It was found that, if the quantity, ${\cal C}$,
defined as the circumference is less than approximately $1.168$, a common
apparent horizon surrounds the multi-black-hole configuration.

The results of all these searches for naked spindle singularities
are controversial but could be resolved if the presence or absence of the EH
could be determined. One could
demonstrate numerically whether or not Wald's view of ST's results is correct
by using existing EH finders \cite{libson96,masso98} in a relevant simulation.
Of course, this could only be effective if the simulation covered enough of the
spacetime to include (part of) the EH.

\subsection{Critical behavior in collapse}
\subsubsection{Gravitational collapse simulations}
For a more detailed discussion of critical behavior see
\cite{gundlach99}. Since Gundlach's {\it Living Review} article has appeared,
the updates in this section will be restricted to results I find especially
interesting.

Critical behavior was originally found by Choptuik
\cite{choptuik93} in a numerical study of the collapse of a spherically
symmetric massless scalar field. For recent reviews see
\cite{gundlach96a,gundlach98}. We note that this is the first completely new
phenomenon in general relativity to be discovered by numerical simulation. In
collapse of a scalar field, essentially two things can happen: either
a black hole (BH) forms or the scalar waves pass through each other and
disperse. Choptuik discovered that for any 1-parameter set of initial data
labeled by
$p$, there is a critical value $p^{\ast}$ such that $p > p^{\ast}$ yields
a BH. He found
\begin{equation}
\label{chopscale}
M_{BH} \approx C_F \left( p - p^{\ast} \right)^{\gamma}
\end{equation}
where $M_{BH}$ is the mass of the eventual BH. The constant $C_F$ depends
on the parameter of the initial data that is selected but $\gamma \approx .37$
is the
same
for all choices. Furthermore, in terms of logarithmic variables
$\rho = \ln r + \kappa$, $\tau = \ln (T_0^{\ast} - T_0) + \kappa$
($T_0$ is the proper time of an observer at $r = 0$, where $r$ is the radial
coordinate, with
$T_0^{\ast}$ the finite proper time at which the critical evolution concludes
and $\kappa$ is a constant which scales $r$), the waveform $X$ repeats
(echoes)  at intervals of
$\Delta$ in
$\tau$ if
$\rho$ is
rescaled to $\rho - \Delta$, i.e. $X(\rho - \Delta, \tau - \Delta)
\approx X(\rho, \tau)$. The scaling behavior (\ref{chopscale}) demonstrates
that the minimum BH mass (for
bosons) is zero. The critical solution itself is a
counter-example to cosmic censorship (since the
formation of the zero mass BH causes high curvature regions to
become visible at $r = \infty$).  (See, e.g., the discussion in Hirschmann
and Eardley \cite{hirschmann95b}.) The numerical demonstration of this feature
of the critical solution was provided by Hamad\'{e} and Stewart
\cite{hamade96b}. This result caused Hawking to pay off a bet to Preskill
and Thorne \cite{browne97,johnson97}.

\begin{figure}[bth]
\begin{center}
\caption{This figure is the final frame of an animation of Type II
critical behavior in Einstein-Yang-Mills collapse. Note the echoing in
the near-critical solution. For the entire movie and related references
see \protect \cite{choptuik99}.}
\end{center}
\end{figure}

Soon after this discovery, scaling and critical phenomena were found in
a variety of contexts. Abrahams and Evans \cite{abrahams93} discovered the
same phenomenon in axisymmetric gravitational wave collapse with a different
value of $\Delta$ and, to within numerical error, the same value of $\gamma$.
(Note that the rescaling of $r$ with $e^{\Delta} \approx 30$ required
Choptuik to use adaptive mesh refinement (AMR) to distinguish subsequent
echoes. Abrahams and
Evans' smaller $\Delta$ ($e^{\Delta} \approx 1.8$) allowed them to see
echoing with their $2+1$ code without AMR.) 
Gar\-fin\-kle~\cite{garfinkle95} confirmed Choptuik's results with a
completely
different algorithm that does not require AMR. He used Goldwirth and
Piran's \cite{goldwirth87} method of simulating Christodoulou's
\cite{christodoulou87} formulation of the spherically symmetric scalar field in
null
coordinates. This formulation allowed the grid
to be automatically rescaled by choosing the edge of the grid to be the
null ray that just hits the central observer at the end of the critical
evolution. (Missing points of null rays that cross the central observer's
world line are replaced by interpolation between those that remain.)
Hamad\'{e} and Stewart \cite{hamade96b} have also repeated Choptuik's
calculation using null coordinates and AMR. They are able to achieve greater
accuracy and find $\gamma \approx .374$.

\subsubsection{Critical solutions as an eigenvalue problem}
Evans and Coleman \cite{evans94} realized that self-similar rather than
self-periodic collapse might be more tractable both numerically (since
ODE's rather than PDE's are involved) and analytically. They discovered
that a collapsing radiation fluid had that desirable property. (Note that
self-similarity (homothetic motion) is incompatible with AF
\cite{eardley86,garfinkle87}. However, most of the action occurs in the center
so that a match of the self-similar inner region to an outer AF one should
always be possible.) In a series of papers, Hirschmann and
Eard\-ley~\cite{hirschmann95a,hirschmann95b} developed a (numerical)
self-similar solution to the spherically symmetric {\it complex} scalar field
equations. These are ODE's with too many boundary conditions causing a solution
to exist only for certain fixed values of
$\Delta$. Numerical
solution of this eigenvalue problem allows very accurate determination of
$\Delta$. The self-similarity also allows accurate
calculation of $\gamma$ as follows:
The critical $p = p^{\ast}$ solution is unstable to a small change in $p$. At
any time $t$ (where $t < 0$ is
increasing toward zero),
the amplitude $a$ of the perturbation exhibits power law growth:
\begin{equation}
\label{eardley1}
a \propto (p - p^{\ast}) (-t)^{- \kappa}
\end{equation}
where $\kappa > 0$. At any fixed $t$, larger $a$ implies larger $M_{BH}$.
Equivalently, any fixed amplitude $a = \delta$ will be reached faster for
larger eventual $M_{BH}$. Scaling arguments give the dependence of
$M_{BH}$ on the time at which any fixed amplitude is reached:
\begin{equation}
\label{eardley2}
M_{BH} \propto (-t_1)
\end{equation}
where
\begin{equation}
\label{eardley3}
(p - p^{\ast}) (-t_1)^{\kappa} \propto \delta .
\end{equation}
Thus
\begin{equation}
\label{eardley4}
M_{BH} \propto (p - p^{\ast})^{1/\kappa} .
\end{equation}
Therefore, one need only identify the growth rate of the unstable mode to
obtain an accurate value of $\gamma = 1 / \kappa$. It is not necessary to
undertake the entire dynamical evolution or probe the space of initial
data. (While this procedure allowed Hirschmann and Eardley to obtain $\gamma =
0.387106$ for the complex scalar field solution, they later found
\cite{hirschmann97} that, in this regime, the complex scalar field has 3
unstable modes. This means \cite{gundlach97c,gundlach99} that the
Eardley-Hirschmann solution is not a critical solution. A perturbation
analysis indicates that the critical solution for the complex scalar field is
the discretely self-similar one found for the real scalar field
\cite{gundlach97c}.) Koike et al~\cite{koike95} obtain
$\gamma = 0.35580192$ for the Evans-Coleman solution. Although the similarities
among the critical exponents $\gamma$ in the collapse computations suggested a
universal value, Maison
\cite{maison95} used these same scaling-perturbation methods to show that
$\gamma$ depends on the equation of state $p = k \rho$ of the fluid in the
Evans-Coleman solution. Gundlach \cite{gundlach95} used a similar approach
to locate Choptuik's critical solution accurately.
This is much harder due to its discrete self-similarity. He reformulates
the model as nonlinear hyperbolic
boundary value problem with eigenvalue $\Delta$ and finds $\Delta
= 3.4439$.  As with the self-similar solutions described above, the critical
solution is found directly without the need to perform a dynamical evolution
or explore the space of initial data.
Hara et al extended the renormalization group approach of \cite{koike95} and
applied it to the continuously-self-similar case \cite{hara96}. (For an
application of renormalization group methods to cosmology see \cite{iguchi97}.) 

\subsubsection{Recent results}
Gundlach \cite{gundlach97c} completed his eigenvalue analysis of the Choptuik
solution to find the growth rate of the unstable mode to be
$\gamma = .374 \pm .001$. He also predicted a periodic ``wiggle'' in the
Choptuik mass scaling relation. This was later observed numerically by Hod and
Piran \cite{hod97b}. Self-similar critical behavior has been seen in string
theory related axion-dilaton models \cite{eardley95,hamade96a} and in the
nonlinear $\sigma$-model \cite{hirschmann97}. Garfinkle and Duncan have shown
that subcritical collapse of a spherically symmetric scalar field yields a
scaling relation for the maximum curvature observed by the central observer
with critical parameters that would be expected on the basis of those found for
supercritical collapse \cite{garfinkle98}.

Choptuik et al \cite{choptuik96} have generalized the original Einstein-scalar
field calculations to the Einstein-Yang-Mills (EYM) (for $SU(2)$) case. Here
something new was found. Two types of behavior appeared depending on the
initial data. In Type I, BH formation had a non-zero mass threshold. The
critical solution is a regular, unstable solution to the EYM equations found
previously by Bartnik and McKinnon \cite{bartnik88}. In Type II collapse, the
minimum BH mass is zero with the critical solution similar to that of Choptuik
(with a different $\gamma \approx 0.20$, $\Delta \approx 0.74$). Gundlach has
also looked at this case with the same results \cite{gundlach97b}. The Type I
behavior arises when the collapsing system has a metastable static solution in
addition to the Choptuik critical one \cite{gundlach96b}.

Brady, Chambers and Gon\c{c}alves \cite{chambers97b,brady97a} conjectured that
addition of a mass to the scalar field of the original model would break scale
invariance and might yield a distinct critical behavior. They found numerically
the same Type I and II ``phases'' seen in the EYM case. The Type II solution
can be understood as perturbations of Choptuik's original model with a small
scalar field mass $\mu$. Here small means that $\lambda \mu << 1$ where
$\lambda$ is the spatial extent of the original nonzero field region. (The
scalar field is well within the Compton wavelength corresponding to $\mu$.) On
the other hand,
$\lambda \mu >> 1$ yields Type I behavior. The minimum mass critical solution
is an unstable soliton of the type found by Seidel and Suen \cite{seidel94}.
The massive scalar field can be treated as collapsing dust to yield a criterion
for BH formation \cite{goncalves97}.

The Choptuik solution has also been found to be the critical solution for
charged scalar fields \cite{gundlach96b,hod97a}. As $p \to p^*$, $Q/M \to 0$ for
the black hole. $Q$ obeys a power law scaling. Numerical study of the critical
collapse of collisionless matter (Einstein-Vlasov equations) has yielded a
non-zero minimum BH mass \cite{rein98,olabarrieta00}. Bizo\'{n} and Chmaj
\cite{bizon98} have considered the critical collapse of skyrmions.

An astrophysical application of BH critical phenomena has been considered by
Nimeyer and Jedamzik \cite{niemeyer97} and Yokoyama \cite{yokoyama98}. They
consider its implications for primordial BH formation and suggest that it could
be important.

\subsubsection{Going further}
The question is then why these critical phenomena should appear in so many
collapsing gravitational systems. The discrete self-similarity of Choptuik's
solution may be understood as scaling of a limit cycle
\cite{hara96}. (The self-similarity of other systems may be understood as
scaling of a limit point.) Garfinkle
\cite{garfinkle97} originally conjectured that the scale invariance of
Einstein's equations might provide an underlying explanation for the
self-similarity and discrete-self-similarity found in collapse and
proposed a spacetime slicing which might manifestly show this. In fact,
he later showed (with Meyer) \cite{garfinkle98a} that, while the originally
proposed slicing failed, a foliation based on maximal slicing did make the
scaling manifest. These ideas formed the basis of a much more ambitious program
by Garfinkle and Gundlach to use underlying actual or approximate symmetries to
construct coordinate systems for numerical simulations \cite{garfinkle99a}. 

An interesting ``toy model'' for general relativity in many contexts has
been wave maps, also known as nonlinear $\sigma$ models. One of these contexts
is critical collapse \cite{hirschmann97}. Recently and independently, Bizo\'{n}
et al \cite{bizon99} and Liebling et al
\cite{liebling99} evolved wave maps from the base
space of $3 + 1$ Minkowski space to the target space $S^3$. They found critical
behavior separating singular and nonsingular solutions. For some families of
initial data, the critical solution is self-similar and is an intermediate
attractor. For others, a static solution separates the singular and nonsingular
solutions. However, the static solution has several unstable modes and is
therefore not a critical solution in the usual sense. Bizo\'{n} and Tabor
\cite{bizon01} have studied Yang-Mills fields in $D + 1$ dimensions and
found that generic solutions with sufficiently large initial data blow up in
a finite time and that the mechanism for blowup depends on $D$. Husa et al
\cite{husa00} then considered the collapse of
$SU(2)$ nonlinear sigma models coupled to gravity and found a discretely
self-similar critical solution for sufficiently large dimensionless coupling
constant. They also observe that for sufficiently small coupling constant
values, there is a continuously self-similar solution. Interestingly, there is
an intermediate range of coupling constant where the discrete self-similarity
is intermittent \cite{thornburg00}.

Until recently, only Abrahams and Evans \cite{abrahams93} had ventured beyond
spherical symmetry. The first additional departure has been made by Gundlach
\cite{gundlach97a}. He considered spherical and non-spherical perturbations of
$P = \textstyle{{1} \over {3}} \rho$ perfect fluid collapse. Only the original
(spherical) growing mode survived.

Recently, critical phenomena have been explored in $2 + 1$ gravity.
Pretorius and Choptuik \cite{pretorius00} numerically evolved circularly
symmetric scalar field collapse in $2 + 1$ anti-deSitter space. They found a
continuously self-similar critical solution at the threshold for black hole
formation. The BH's which form have BTZ \cite{banados92} exteriors with
strong curvature, spacelike singularities in the interior. Remarkably,
Garfinkle obtained an analytic critical solution by assuming continuous
self-similarity which agrees quite well with the one obtained numerically
\cite{garfinkle00}.

\subsection{Nature of the singularity in charged or rotating black holes}
\subsubsection{Overview}
Unlike the simple singularity structure of the Schwarzschild solution, where
the event horizon
encloses a spacelike singularity at $r=0$, charged and/or rotating BH's
have a much richer singularity structure. The extended spacetimes have an inner
Cauchy horizon (CH) which is the boundary of predictability. To the future
of the CH lies a timelike (ring) singularity \cite{wald84}.
Poisson and Israel \cite{poisson89,poisson90} began an analytic study of
the effect of perturbations on the CH. Their goal was to check conjectures
that the blue-shifted
infalling radiation during collapse would convert the CH into a true
singularity and thus prevent an observer's passage into the rest of the
extended regions. By including both ingoing and back-scattered
outgoing radiation, they
find for the Reissner-Nordstrom (RN) solution that the mass function
(qualitatively
$R_{\alpha \beta \gamma \delta} \propto M / r^3$) diverges at the CH
(mass inflation). However, Ori showed both for RN and Kerr
\cite{ori91,ori92} that the metric perturbations are finite (even though
$R_{\mu \nu \rho \sigma}R^{\mu \nu \rho \sigma}$ diverges) so that an
observer would not be destroyed by tidal forces
(the tidal distortion would be finite)
and could survive passage through the CH. A numerical solution of the
Einstein-Maxwell-scalar field equations would test these perturbative
results.

For an excellent, brief review of the history of this topic see the
introduction in \cite{ori99b}.

\subsubsection{Numerical studies}
Gnedin and Gnedin \cite{gnedin93} have numerically evolved the
spherically symmetric Einstein-Maxwell with massless scalar field equations
in a $2+2$ formulation. The initial conditions place a scalar field on part of
the
RN event horizon (with zero field on the rest). An asymptotically null or
spacelike singularity whose shape depends on the strength of the initial
perturbation replaces the CH. For a sufficiently strong perturbation, the
singularity is Schwarzschild-like. Although they claim to have found that
the CH evolved to become a spacelike singularity, the diagrams in
their paper show at least part of the final singularity to be null or
asymptotically
null in most cases.

Brady and Smith \cite{brady95b} used the Goldwirth-Piran formulation
\cite{goldwirth87} to study the same problem. They assume the spacetime
is RN for $v < v_0$. They follow the evolution of the CH
into a null singularity, demonstrate mass inflation, and support (with
observed exponential decay of the metric component $g$) the validity of
previous analytic results \cite{poisson89,poisson90,ori91,ori92} including the
``weak'' nature of the singularity that forms. They find that the
observer hits the null CH singularity before falling into the curvature
singularity at $r = 0$. Whether or not these results are in conflict with
Gnedin and Gnedin \cite{gnedin93} is unclear \cite{bonanno95}.
However, it has become clear that Brady and Smith's conclusions are correct.
The collapse of a scalar field in a charged, spherically symmetric spacetime
causes an initial RN CH to become a null singularity except at $r = 0$ where it
is spacelike. The observer falling into the BH experiences (and potentially
survives) the weak, null singularity \cite{ori91,ori92,brady95a} before the
spacelike singularity forms. This has been confirmed by Droz \cite{droz96}
using a plane wave model of the interior and by Burko \cite{burko97a} using a
collapsing scalar field. See also \cite{burko98,burko97e}.

Numerical studies of the interiors of non-Abelian black holes have
been carried out by Breitenlohner et al
\cite{breitenlohner97a,breitenlohner97b} and by Gal'tsov et al
\cite{donets97,galtsov97a,galtsov97b,galtsov97c} (see also
\cite{zotov97}). Although there appear to be conflicts between the two groups,
the differences can be attributed to misunderstandings of each other's notation
\cite{breitenlohner97c}. The main results include an interesting oscillatory
behavior of the metric.

\begin{figure}[bth]
\begin{center}
\caption{Figure 1 from \protect \cite{burko98a} is a schematic diagram of
the singularity structure within a spherical charged black hole.}
\end{center}
\end{figure}

The current status of the topic of singularities within BH's includes an
apparent conflict between the belief
\cite{belinskii71a} and numerical evidence \cite{berger98c} that the generic
singularity is strong, oscillatory, and spacelike and analytic evidence that
the singularity inside a generic (rotating) BH is weak, oscillatory (but in a
different way), and null
\cite{ori99a}. See the discussion at the end of \cite{ori99a}. 

Various recent perturbative results reinforce the belief that the
singularity within a ``realistic'' (i.e. one which results from
collapse) black hole is of the weak, null type described by Ori
\cite{ori91,ori92}. Brady et al \cite{brady98a} performed an analysis
in the spirit of Belinskii et al \cite{belinskii71a} to argue that the
singularity is of this type. They constructed an asymptotic expansion
about the CH of a black hole formed in gravitational collapse without
assuming any symmetry of the perturbed solution. To illustrate their
techniques, they also considered a simplified ``almost'' plane
symmetric model. Actual plane symmetric models with weak, null
singularities were constructed by Ori \cite{ori98}. 

The best numerical evidence for the nature of the singularity in realistic
collapse is Hod and Piran's simulation of the gravitational collapse of a
spherically symmetric, charged scalar field \cite{hod98,hod99}. Rather than
start with (part of) a RN spacetime which already has a singularity (as in,
e.g., \cite{brady95b}), they begin with a regular spacetime and follow its
dynamical evolution. They observe mass inflation, the formation of a null
singularity, and the eventual formation of a spacelike singularity. Ori argues
\cite{ori99a} that the rotating black hole case is different and that the
spacelike singularity will never form. No numerical studies beyond perturbation
theory have yet been made for the rotating BH. 

Some insight into the conflict between the cosmological results and those from
BH interiors may be found by comparing the approach to the singularity in Gowdy
\cite{gowdy71} spatially inhomogeneous cosmologies (see Sec.~\ref{gowdy}) with
$T^3$ \cite{berger97b} and $S^2 \times S^1$
\cite{garfinkle99} spatial topologies. Early in the collapse, the boundary
conditions associated with the $S^2 \times S^1$ topology influence the
gravitational waveforms. Eventually, however, the local behavior of the two
spacetimes becomes qualitatively indistinguishable and characteristic of a
(non-oscillatory in this case) spacelike singularity. This may be relevant
because the BH environment imposes effective boundary conditions on the metric
just as topology does.  Unfortunately, no systematic study of the relationship
between the cosmological and BH interior results yet exists.

\subsubsection{Going further}
Replacing the AF boundary conditions with Schwarzchild-deSitter and RN-deSitter
BH's was long believed to yield a counterexample to strong cosmic censorship.
(See
\cite{mellor90,mellor92,poisson97,chambers97a} and references therein for
background and extended discussions.) The stability of the CH is related to the
decay tails of the radiating scalar field. Numerical studies have determined
these to be exponential \cite{brady97c,chambers97a,chambers97c} rather than
power law as in AF spacetimes \cite{burko97c}. The decay tails of the radiation
are necessary initial data for numerical study of CH stability \cite{brady95b}
and are crucial to the development of the null singularity. Analytic studies
had indicated that the CH is stable in RN-deSitter BH's for a restricted range
of parameters near extremality. However, Brady et al \cite{brady98} have shown
(using linear perturbation theory) that, if one includes the backscattering of
outgoing modes which originate near the event horizon, the CH is always unstable
for all ranges of parameters. Thus RN-de Sitter BH's appear not to be a
counterexample to strong cosmic censorship. Numerical studies are needed to
demonstrate the existence of a null singularity at the CH in nonlinear
evolution.

Extension of these studies to AF rotating BH's has yielded the surprising
result that the tails are not necessarily power law and differ for different
fields. Frame dragging effects appear to be responsible \cite{hod00}.

As a potentially useful approach to the numerical study of singularities,we
consider H\"{u}bner's
\cite{huebner96a,huebner96b,huebner98} numerical scheme to evolve on a conformal
compactified grid using Friedrich's formalism \cite{Friedrich88}. He considers
the spherically symmetric scalar field model in a $2+2$ formulation. So far,
this code has been used in this context to locate singularities.  It was also
used to search for Choptuik's scaling \cite{choptuik93} and failed to produce
agreement with Choptuik's results \cite{huebner96a}. This was probably due to
limitations of the code rather than inherent problems with the conformal method.

\section{Singularities in cosmological models}
\subsection{Singularities in spatially homogeneous cosmologies}
The generic singularity in spatially homogeneous cosmologies is reasonably well
understood. The approach to it asymptotically falls into two classes. The
first, called asymptotically velocity term dominated (AVTD)
\cite{eardley72,isenberg90}, refers to a cosmology that approaches the
Kasner (vacuum, Bianchi I) solution \cite{kasner25} as $\tau \to
\infty$. (Spatially homogeneous universes can be described as a sequence of
homogeneous spaces labeled by
$\tau$. Here we shall choose $\tau$ so that $\tau = \infty$ coincides with the
singularity.) An example of such a solution is the vacuum Bianchi II model
\cite{taub51} which begins with a fixed set of Kasner-like anisotropic
expansion rates, and, possibly, makes one change of the rates in a prescribed
way (Mixmaster-like bounce) and then continues to $\tau = \infty$ as a fixed
Kasner solution. In contrast are the homogeneous cosmologies which display
Mixmaster dynamics such as vacuum Bianchi VIII and IX
\cite{belinskii71b,misner69,halpern87} and Bianchi VI$_0$ and Bianchi I with
a magnetic field \cite{leblanc95,berger96a,leblanc97}. Jantzen
\cite{jantzen86} has discussed other examples. Mixmaster dynamics describes an
approach to the singularity which is a sequence of Kasner epochs with a
prescription, originally due to Belinskii, Khalatnikov, and Lifshitz (BKL)
\cite{belinskii71b}, for relating one Kasner epoch to the next. Some of the
Mixmaster bounces (era changes) display sensitivity to initial conditions one
usually associates with chaos and in fact Mixmaster dynamics is chaotic
\cite{cornish97b,motter01}. Note that we consider an {\it epoch} to be a
subunit of an {\it era}. In some of the literature, this usage is reversed. The
vacuum Bianchi I (Kasner) solution is distinguished from the other Bianchi
types in that the spatial scalar curvature
$^3\!R$, (proportional to) the minisuperspace (MSS) potential
\cite{misner69,ryan75}, vanishes identically. But
$^3\!R$ arises in other Bianchi types due to spatial dependence of the metric
in a coordinate basis. Thus an AVTD singularity is also characterized as a
regime in which terms containing or arising from spatial derivatives no longer
influence the dynamics. This means that the Mixmaster models do not have an
AVTD singularity since the influence of the spatial derivatives (through the
MSS potential) never disappears---there is no last bounce.  A more general
review of numerical cosmology has been given by Anninos \cite{anninos01}.

\subsection{Numerical Methods}

\subsubsection{Symplectic Methods}
Symplectic numerical methods have proven useful in studies of the approach to
the singularity in cosmological models \cite{berger97c}. Symplectic ODE and PDE
\cite{fleck76,moncrief83} methods comprise a type of operator splitting.  An
outline of the method (for one degree of freedom) follows. Details of the
application to the Gowdy model (PDE's in one space and one time direction) are
given elsewhere \cite{berger93}.

For a field $q(t)$ and its conjugate momentum $p(t)$ split
the Hamiltonian operator into kinetic and potential energy subhamiltonians.
Thus, for an arbitrary potential $V(q)$,
\begin{equation}
H= \textstyle{1 \over 2}p^2+V(q)=H_1(p)+H_2(q) .
\end{equation}
If the vector $X = (p,q)$
defines the variables at time $t$, then the time evolution is given by
\begin{equation}
{{dX} \over {dt}}=\{H,X\}_{PB}\equiv AX
\end{equation}
where $\{ \quad \}_{PB}$ is the Poisson bracket. The usual exponentiation
yields an evolution operator
\begin{equation}
\label{evapprox}
e^{A\Delta t}=e^{A_1(\Delta t/ 2)}e^{A_2\Delta
t}e^{A_1(\Delta t/2)}+O(\Delta t^3)
\end{equation}
for $A = A_1 + A_2$ the generator of the time evolution.
Higher order accuracy may be obtained by a better
approximation to the evolution operator \cite{suzuki90,suzuki91}.
This method is useful when exact solutions for the
subhamiltonians are known. For the given $H$, variation of $H_1$ yields the
solution
\begin{equation}
q=q_0+p _0\,\Delta t\quad,\quad p =p_0,
\end{equation}
while that of $H_2$ yields
\begin{equation}
q=q_0\quad,\quad p =p_0-\left.{{d V}
\over {d q}}\right|_{q_0}\Delta t \ \ .
\end{equation}
Note that $H_2$ is exactly solvable for any potential $V$ no matter how
complicated although the required
differenced form of the potential gradient may be non-trivial.
One evolves from $t$  to  $t + \Delta t$ using the exact
solutions to the subhamiltonians according to the
prescription given by the approximate evolution operator (\ref{evapprox}).
Extension to more degrees of freedom and to fields is straightforward
\cite{berger93,berger96b}.

\subsubsection{Other Methods}
Symplectic methods can achieve convergence far beyond that of their formal
accuracy if the full solution is very close to the exact solution from one of
the subhamiltonians. Examples where this is the case are given in
\cite{berger96c,berger97b}. On the other hand, because symplectic algorithms are
a type of operator splitting, suboperators can be subject
to instabilities that are suppressed by the full operator. An example of this
may be found in \cite{berger01}. Other types of PDE solvers are more effective
for such spacetimes. One currently popular method is iterative Crank-Nicholson
(see \cite{teukolsky00}) which is, in effect, an implicit method without matrix
inversion. It was first applied to numerical cosmology by Garfinkle
\cite{garfinkle99} and was recently used in this context to evolve $T^2$
symmetric cosmologies
\cite{berger01}. 

As pointed out in \cite{berger93,berger97b,berger01,berger98a}, spiky features
in collapsing inhomogeneous cosmologies will cause any fixed spatial resolution
to become inadequate. Such evolutions are therefore candidates for adaptive mesh
refinement such as that implemented by Hern and Stuart \cite{hern97,hern00}.

\subsection{Mixmaster dynamics}
\subsubsection{Overview}
Belinskii, Khalatnikov, and Lifshitz \cite{belinskii71b} (BKL) described the
singularity approach of
vacuum Bianchi IX cosmologies as an infinite sequence of
Kasner \cite{kasner25} epochs whose indices change when the scalar curvature
terms in Einstein's equations become important. They were able to describe the
dynamics approximately by a map evolving a discrete set of parameters from one
Kasner epoch to the next \cite{belinskii71b,chernoff83}. For example, the Kasner
indices for the power law dependence of the anisotropic scale factors can be
parametrized by a
single variable $u \ge 1$. BKL determined that
\begin{equation}
\label{umap}
u_{n+1}=\left\{ \matrix{u_n-1 \quad \quad,\quad 2\le u_n\hfill\cr
  (u_n-1)^{-1}\quad,\quad1\le u_n\le 2\hfill\cr} \right. \quad .
\end{equation}
The subtraction in the denominator
for $1 \le u_n \le 2$ yields the sensitivity to initial conditions
associated with Mixmaster dynamics (MD).
Misner \cite{misner69} described the same
behavior in terms of the model's volume and anisotropic
shears. A multiple of the scalar curvature
acts as an outward moving potential in
the anisotropy plane.  Kasner epochs become straight line
trajectories moving outward along a potential corner while bouncing from one side
to the other.
A change of corner ends a BKL era when $u \to (u-1)^{-1}$.
Numerical evolution of Einstein's equations was used to
explore the accuracy of the BKL map as a descriptor of the dynamics as well as
the implications of the map \cite{moser73,rugh90a,rugh90b,berger94}. Rendall has
studied analytically the validity of the BKL map as an approximation to the true
trajectories \cite{rendall97b}.

Later, the BKL sensitivity to initial conditions was
discussed in the language of chaos \cite{barrow82,khalatnikov85}. An extended
application of Bernoulli shifts and Farey trees was given by Rugh \cite{rugh94}
and repeated by Cornish and Levin \cite{cornish97a}. However, the chaotic nature
of Mixmaster dynamics was questioned when numerical evolution of the Mixmaster
equations yielded zero Lyapunov exponents (LE's)
\cite{francisco88,burd90,hobill91}. (The LE measures the divergence of
initially nearby trajectories. Only an exponential divergence, characteristic
of a chaotic system, will yield positive exponent.) Other numerical studies
yielded positive LE \cite{pullin91}. This issue was resolved when the LE was
shown numerically and analytically to depend on the choice of time variable
\cite{rugh90a,berger91,ferraz91}. Although MD itself is well-understood, its
characterization as chaotic or not had been quite controversial
\cite{hobill94}. 

LeBlanc et al \cite{leblanc95} have shown (analytically and numerically)
that MD can arise in Bianchi VI$_0$
models with magnetic fields (see also \cite{ma94}). In essence, the magnetic
field provides the wall needed to close the potential in a way that yields the
BKL map for $u$ 
\cite{berger96a}. A similar study of magnetic Bianchi I has been given by
LeBlanc \cite{leblanc97}. Jantzen has discussed which vacuum and electromagnetic
cosmologies could display MD \cite{jantzen86}.

Cornish and Levin (CL) \cite{cornish97b} identified a coordinate
invariant way to characterize MD. Sensitivity to initial conditions can lead to
qualitatively distinct outcomes from initially nearby trajectories. While the
LE measures the exponential divergence of the trajectories, one can also
``color code'' the regions of initial data space corresponding to particular
outcomes. A chaotic system will exhibit a fractal pattern in the colors. CL
defined the following set of discrete outcomes: During a numerical evolution,
the BKL parameter $u$ is evaluated from the trajectories. The first time $u >
7$ appears in an approximately Kasner epoch, the trajectory is examined to see
which metric scale factor has the largest time derivative. This defines three
outcomes and thus three colors for initial data space. However, one can easily
invent prescriptions other than that given by Cornish and Levin
\cite{cornish97b} which would yield discrete outcomes. The fractal nature of
initial data space should be common to all of them. It is not clear how the
value of the fractal dimension as measured by Cornish and Levin would be
affected. The CL prescription has been criticized because it requires only the
early part of a trajectory for implementation \cite{motter01}. Actually, this
is the greatest strength of the prescription for numerical work. It replaces a
single representative, infinitely long trajectory by (easier to compute)
arbitrarily many trajectory fragments. 

\begin{figure}[bth]
\begin{center}
\caption{The algorithm of \protect \cite{berger96c} is used to generate a
Mixmaster trajectory with more than 250 bounces. The trajectory is shown
in the rescaled anisotropy plane with axes $\beta_\pm/|\Omega|$. The
rescaling fixes (asymptotically) the location of the bounces.}
\end{center}
\end{figure}

To study the CL fractal and ergodic properties of Mixmaster evolution
\cite{cornish97b}, one could take advantage of a new numerical algorithm due to
Berger, Garfinkle, and Strasser \cite{berger96c}. Symplectic methods are used to
allow the known exact solution for a single Mixmaster bounce \cite{taub51} to be
used in the ODE solver. Standard ODE solvers \cite{press92} can take large time
steps in the Kasner segments but must slow down at the bounce. The new algorithm
patches together exactly solved bounces. Tens of orders of magnitude
improvement in speed are obtained while the accuracy of machine precision is
maintained. In
\cite{berger96c}, the new algorithm was used to distinguish Bianchi IX and
magnetic Bianchi VI$_0$ bounces. This required an improvement of the BKL map
(for parameters other than
$u$) to take into account details of the exponential potential.

So far, most recent effort in MD has focused on diagonal (in the frame of the
$SU(2)$ 1-forms) Bianchi IX models. Long ago, Ryan \cite{ryan71} showed that
off-diagonal metric components can contribute additional MSS potentials (e.g. a
centrifugal wall). This has been further elaborated by Jantzen \cite{jantzen01}.

\subsubsection{Recent developments}
The most interesting recent developments in spatially homogeneous Mixmaster
models have been mathematical. Despite the strong numerical evidence that
Bianchi IX, etc. models are well-approximated by the BKL map sufficiently close
to the singularity (see, e.g., \cite{berger96c}), there was very little rigorous
information on the nature of these solutions. Recently, the existence of a
strong singularity (curvature blowup) was proved for Bianchi VIII and IX
collapse by Ringstr\"{o}m \cite{ringstrom99,ringstrom00} and for magnetic
Bianchi VI$_0$ by Weaver \cite{weaver99b}. A remaining open question is how
closely an actual Mixmaster evolution is approximated by a single BKL sequence
\cite{rendall97b,ringstrom00}.  Since the Berger et al algorithm
\cite{berger96c} achieves machine level accuracy, it can be used to collect
numerical evidence on this topic. For example, it has been shown that a given
Mixmaster trajectory ceases to track the corresponding sequence of integers
obtained from the BKL map (\ref{umap}) at the point where there have been
enough era-ending (mixing) bounces to lose all the information encoded in
finite precision initial data
\cite{berger96c}. 

\subsubsection{Going further}

There are also numerical studies of Mixmaster dynamics in other
theories of gravity. For example, Carretero-Gonzalez et al
\cite{carretero94} find
evidence of chaotic behavior in Bianchi IX-Brans-Dicke solutions while
Cotsakis et al \cite{cotsakis93} have shown that
Bianchi IX models in 4th order gravity theories
have stable non-chaotic solutions. Barrow and Levin find evidence of chaos in
Bianchi IX Einstein-Yang-Mills (EYM) cosmologies \cite{barrow97}. Their analysis
may be applicable to the corresponding EYM black hole interior solutions.
Bianchi I models with string-theory-inspired matter fields have been found by
Damour and Henneau \cite{damour00b} to have an oscillatory singularity. This is
interesting because many other examples exist where matter fields and/or higher
dimensions suppress such oscillations (see e.g.~\cite{belinskii73}).  Recently,
Coley has considered Bianchi IX brane-world models and found them not to be
chaotic \cite{coley01}.

Finally, we remark on a successful application of numerical Regge calculus in
$3 + 1$ dimensions. Gentle and Miller have been able to evolve the Kasner
solution \cite{gentle97}.

\subsection{Inhomogeneous Cosmologies}
\subsubsection{Overview}
BKL have conjectured that one should expect a generic singularity in spatially
inhomogeneous cosmologies to be locally of the Mixmaster type (local Mixmaster
dynamics (LMD))
\cite{belinskii71b}.  For a review of homogeneous cosmologies,
inhomogeneous cosmologies, and the relation between them, see
\cite{maccallum79}. The main difficulty with the acceptance of this conjecture
has been the controversy over whether the required time slicing can be
constructed globally
\cite{barrow79,grubisic94a}. Montani \cite{montani95},
Belinskii {\cite{belinskii92}, and Kir\-il\-lov and Kochnev
\cite{kirillov87,kirillov93} have pointed out that if the BKL conjecture is
correct, the spatial structure of the singularity could become extremely
complicated as bounces occur at different locations at different times.  BKL
seem to imply \cite{belinskii71b} that LMD should only be expected to occur in
completely general spacetimes with no spatial symmetries. However, LMD is
actually possible in any spatially inhomogeneous cosmology with a local MSS
with a ``closed'' potential (in the sense of the standard triangular potentials
of Bianchi VIII and IX). This closure may be provided by spatial curvature,
matter fields, or rotation. A class of cosmological models which appear to have
local MD are vacuum universes on $T^3 \times R$ with a $U(1)$ symmetry
\cite{moncrief86}. Even simpler plane symmetric
Gowdy cosmologies \cite{gowdy71,berger74} have ``open'' local MSS potentials.
However, these models are interesting in their own right since they have been
conjectured to possess an AVTD singularity \cite{grubisic93}. One way to obtain
these Gowdy models is to allow spatial dependence in one direction in Bianchi I
homogeneous cosmologies \cite{berger74}. It is well-known that addition of
matter terms to homogeneous Bianchi I, Bianchi VI$_0$, and other AVTD models
can yield Mixmaster behavior
\cite{jantzen86,leblanc95,leblanc97}. Allowing spatial dependence in one
direction in such models might then yield a spacetime with LMD. Application of
this procedure to magnetic Bianchi VI$_0$ models yields magnetic Gowdy models
\cite{weaver98,weaver99a}. Of course, Gowdy cosmologies are not the
most general $T^2$ symmetric vacuum spacetimes
\cite{gowdy71,chrusciel90a,berger97g}. Restoring the ``twists'' introduces a
centrifugal wall to close the MSS. Magnetic Gowdy and general $T^2$ symmetric
models appear to admit LMD
\cite{weaver99a,weaver99b,berger01}.

The past few years have seen the development of strong numerical evidence in
support of the BKL claims \cite{berger98c}. The Method of Consistent Potentials
(MCP) \cite{grubisic93} has been used to organize the data obtained in
simulations of spatially inhomogeneous cosmologies
\cite{berger93,berger97b,weaver98,berger98a,berger98c,berger99a,berger01}. The
main idea is to obtain a Kasner-like velocity term dominated (VTD) solution at
every spatial point by solving Einstein's equations truncated by removing
all terms containing spatial derivatives. If the spacetime really is AVTD, all
the neglected terms will be subdominant (exponentially small in variables where
the VTD solution is linear in the time $\tau$) when the VTD solution is
substituted back into them. For the MCP to successfully predict
whether or not the spacetime is AVTD, the dynamics of the full solution must be
dominated at (almost) every spatial point by the VTD solution behavior.
Surprisingly, MCP predictions have proved valid in numerical simulations of
cosmological spacetimes with one \cite{berger98a} and two
\cite{berger93,berger97b,berger01} spatial symmetries. In the case of $U(1)$
symmetric models, a comparison between the observed behavior \cite{berger98a}
and that in a vacuum, diagonal Bianchi IX model written in terms of $U(1)$
variables provides strong support for LMD \cite{berger99b} since the
phenomenology of the inhomogeneous cosmologies can be reproduced by this
rewriting of the standard Bianchi IX MD. 

Polarized plane symmetric cosmologies have been
evolved numerically using standard techniques by Anninos, Centrella, and
Matzner \cite{anninos91a,anninos91b}. The long-term project involving Berger,
Garfinkle, and Moncrief and their students to study the generic cosmological
singularity numerically has been reviewed in detail elsewhere
\cite{berger97c,berger98c,berger01a} but will be discussed briefly here.  

\subsubsection{Gowdy Cosmologies and Their Generalizations \label{gowdy}}
The Gowdy model on $T^3 \times R$ is described by gravitational wave amplitudes
$P(\theta,\tau)$ and $Q(\theta,\tau)$ which propagate in a spatially
inhomogeneous background universe described by $\lambda(\theta,\tau)$. (We note
that the physical behavior of a Gowdy spacetime can be computed from the effect
of the metric evolution on a test cylinder \cite{berger95b}.) We
impose $0 \le \theta \le 2\pi$ and periodic boundary conditions. The time
variable $\tau$ measures the area in the symmetry plane with $\tau = \infty$ a
curvature singularity. 

Einstein's equations split into two groups. The first is
nonlinearly coupled wave equations for dynamical variables $P$ and $Q$ (where
$,_a = \partial / {\partial a}$) obtained from the variation of
\cite{moncrief81}
\begin{eqnarray}
\label{ber-gowdywaveh}
H&=&{\textstyle{1 \over 2}}\int\limits_0^{2\pi } {d\theta 
\,\left[ {\pi _P^2+\kern 1pt\,e^{-2P}\pi _Q^2} \right]}\nonumber \\
  &+&{\textstyle{1 \over 2}}\int\limits_0^{2\pi } {d\theta \,\left[ {e^{-
2\tau }\left( {P,_\theta ^2+\;e^{2P}Q,_\theta ^2} \right)} 
\right]}=H_1+H_2 \  
\end{eqnarray}
where $\pi_P$ and $\pi_Q$ are canonically conjugate to $P$ and $Q$ respectively.
This Hamiltonian has the form required by the symplectic scheme. If the model
is, in fact, AVTD, the approximation in the symplectic numerical scheme should
become more accurate as the singularity is approached. 
The second group of Einstein equations contains the Hamiltonian and
$\theta$-momentum constraints respectively.
These can be expressed as first order equations for $\lambda$ in terms of
$P$ and $Q$.
This break into dynamical and constraint equations removes two of the
most problematical areas of numerical relativity from this model---the initial
value problem and numerical preservation of the constraints. 

For the special
case of the polarized Gowdy model ($Q=0$), $P$ satisfies a linear wave equation
whose exact solution is well-known \cite{berger74}. For this case, it has been
proven that the singularity is AVTD \cite{isenberg90}. This has also been
conjectured to be true for generic Gowdy models \cite{grubisic93}. 

AVTD behavior is defined in
\cite{isenberg90} as follows: Solve the Gowdy wave equations neglecting all
terms containing spatial derivatives. This yields the VTD solution
\cite{berger93}. If the approach to the singularity is AVTD, the full solution
comes arbitrarily close to a VTD solution at each spatial point as $\tau \to
\infty$. As $\tau
\to
\infty$, the VTD solution becomes
\begin{equation}
P(\theta, \tau) \to v(\theta) \tau \quad \quad , \quad \quad Q(\theta, \tau)
\to Q_0(\theta)
\end{equation}
where $v > 0$. Substitution in the wave equations shows that this behavior is
consistent with asymptotic exponential decay of all terms containing spatial
derivatives only if $0 \le v < 1$ \cite{grubisic93}. We have shown that, except
at isolated spatial points, the nonlinear terms in the wave equation for $P$
drive
$v$ into this range \cite{berger97b,berger97c}. The exceptional points occur
when coefficients of the nonlinear terms vanish and are responsible for the
growth of spiky features seen in the wave forms \cite{berger93,berger97b}. We
conclude that generic Gowdy cosmologies have an AVTD singularity except at
isolated spatial points
\cite{berger97b,berger97c}. This has been confirmed by Hern and Stuart
\cite{hern97} and by van Putten \cite{vanputten97}. After the nature of the
solutions became clear through numerical experiments, it became possible to use
Fuchsian asymptotic methods to prove that Gowdy solutions with
$0 < v < 1$ and AVTD behavior almost everywhere are generic
\cite{kichenassamy98}. These methods have been recently been applied to Gowdy
spacetimes with $S^2 \times S^1$ and $S^3$ topologies with similar conclusions
\cite{stahl01}.

One striking property of the Gowdy models are the development of ``spiky
features'' at isolated spatial points where the coefficient of a local
``potential term'' vanishes \cite{berger93,berger97b}. Recently, Rendall and
Weaver have shown analytically how to generate such spikes from a Gowdy
solution without spikes
\cite{rendall01}.

Addition of a magnetic field to the vacuum Gowdy models (plus a topology change)
which yields the inhomogeneous generalization of magnetic Bianchi VI$_0$ models
provides an additional potential which grows exponentially if $0 < v < 1$.
Local Mixmaster behavior has recently been observed in these magnetic
Gowdy models \cite{weaver98,weaver99a}.

Garfinkle has used a vacuum Gowdy model with $S^2 \times S^1$ spatial topology
to test an algorithm for axis regularity \cite{garfinkle99}. Along the way, he
has shown that these models are also AVTD with behavior at generic spatial
points that is eventually identical to that in the $T^3$ case. Comparison of
the two models illustrates that topology or other global or boundary
conditions are important early in the simulation but become irrelevant as the
singularity is approached.

Gowdy spacetimes are not the most general $T^2$ symmetric vacuum cosmologies.
Certain off-diagonal metric components (the twists which are $g_{\theta
\sigma}$, $g_{\theta \delta}$ in the notation of (\ref{ber-gowdywaveh})) have
been set to zero \cite{gowdy71}. Restoring these terms (see
\cite{chrusciel90,berger97g}) yields spacetimes that are not AVTD but rather
appear to exhibit a novel type of LMD \cite{berger01,weaver01}. The LMD in these
models is an inhomogeneous generalization of non-diagonal Bianchi models with
``centrifugal'' MSS potential walls \cite{ryan75,jantzen01} in
addition to the usual curvature walls. In \cite{berger01},
remarkable quantitative agreement is found between predictions of the MCP and
numerical simulation of the full Einstein equations. A version of the code
with AMR has been developed \cite{belanger01}.  (Asymptotic methods have been
used to prove that the polarized version of these spacetimes have AVTD solutions
\cite{isenberg98}.)

\subsubsection{$U(1)$ Symmetric Cosmologies}
Moncrief has shown \cite{moncrief86} that cosmological models on $T^3 \times R$
with a spatial $U(1)$ symmetry\index{U(1) symmetric cosmologies}
can be described by five degrees of freedom
$\{ x,z, \Lambda, \varphi, \omega \}$ and their respective conjugate momenta
$\{ p_x, p_z, p_{\Lambda}, p, r \}$.  All variables are functions of spatial
variables $u$, $v$ and time, $\tau$. Einstein's
equations can be obtained by variation of
\begin{eqnarray}
\label{Hu1}
H &=& \oint \oint du \kern 1pt dv \,{\cal H} \nonumber \\
\nonumber \\
&=& \oint \oint du \kern 1pt dv \left( {\textstyle{1 \over 8}}p_z^2\,+\,
{\textstyle {1
\over 2}} e^{4z}p_x^2\,+\,{\textstyle{1 \over 8}}p^2\,+\,{\textstyle{1 \over
2}}e^{4\varphi }r^2\,-\,{\textstyle{1
\over 2}}p_\Lambda ^2+2p_\Lambda  \right) \nonumber \\
\nonumber \\
&& \,+\,e^{-2\tau } \oint \oint du \kern 1pt 
dv \left\{  \left( {e^\Lambda e^{ab}} \right) ,_{ab}\,-\, \left( {e^\Lambda
e^{ab}} \right) ,_a\Lambda ,_b \,\right. \nonumber \\
&&+\,e^\Lambda   \left[ 
\left( {e^{-2z}}
\right) ,_u x,_v\,-\, \left( {e^{-2z}} \right) ,_v x,_u \right] \nonumber \\
\nonumber \\
&& \left. +\,2e^\Lambda e^{ab}\varphi ,_a\varphi ,_b\,+\,{\textstyle{1 \over 2}}
e^\Lambda e^{-4\varphi }e^{ab}\omega ,_a\omega ,_b \right\} \nonumber \\
\nonumber \\
&=& H_1+H_2 \ .
\end{eqnarray}
Here $\varphi$ and $\omega$ are analogous to $P$ and $Q$ while $e^\Lambda$
is a conformal factor for the metric $e_{ab}(x,z)$ in the $u$-$v$ plane
perpendicular to the symmetry direction.  Symplectic methods are still
easily applicable. Note particularly that
$H_1$ contains two copies of the Gowdy $H_1$ plus a free particle term and is
thus exactly solvable. The potential term
$H_2$ is very complicated. However, it still contains no momenta so its
equations are trivially exactly solvable.  However, issues of spatial
differencing are problematic. (Currently, a scheme due to Norton
\cite{norton92} is used. A spectral evaluation of derivatives \cite{finn97}
which has been shown to work in Gowdy simulations \cite{berger97f} does not
appear to be helpful in the
$U(1)$ case.) A particular solution to the initial value problem is used since
the general solution is not available
\cite{berger97c}. 

\begin{figure}[bth]
\begin{center}
\caption{Behavior of the gravitational wave amplitude at a typical
spatial point in a collapsing $U(1)$ symmetric cosmology. For details see
\protect \cite{berger98a,berger98c}.}
\end{center}
\end{figure}

Current limitations of the $U(1)$ code do not affect simulations for the
polarized case since problematic spiky features do not develop. Polarized
models have $r = \omega = 0$. Grubi\u{s}i\'{c} and Moncrief
\cite{grubisic94} have conjectured that these polarized models are AVTD. The
numerical simulations provide strong support for this conjecture
\cite{berger97c,berger97e}. Asymptotic methods have been used to prove that an
open set of AVTD solutions exist for this case \cite{isenberg00}.

\subsubsection{Going further}
The MCP indicates that the term
containing gradients of $\omega$ in (\ref{Hu1}) acts as a Mixmaster-like
potential to drive the system away from AVTD behavior in generic $U(1)$ models
\cite{berger98b}.  Numerical simulations provide support for this suggestion
\cite{berger97c,berger98a}.  Whether this potential term grows or decays depends
on a function of the field momenta. This in turn is restricted by the Hamiltonian
constraint. However, failure to enforce the constraints can cause an erroneous
relationship among the momenta to yield qualitatively wrong behavior. There is
numerical evidence that this error tends to suppress Mixmaster-like behavior
leading to apparent AVTD behavior in extended spatial regions of $U(1)$
symmetric cosmologies
\cite{berger96b,berger97a}. In fact, it has been found 
\cite{berger98a}, that when the Hamiltonian constraint is enforced at every
time step, the predicted local oscillatory behavior of the approach to the
singularity is observed. (The momentum constraint is not enforced.) (Note that
in a numerical study of vacuum Bianchi IX homogeneous cosmologies, Zardecki
obtained a spurious enhancement of Mixmaster oscillations due to constraint
violation \cite{zardecki83,hobill91}. In this case, the constraint violation
introduced negative energy.)

Mixmaster simulations with the new algorithm \cite{berger96c} can easily evolve
more than 250 bounces reaching $|\Omega| \approx 10^{62}$. This compares to
earlier simulations yielding 30 or so bounces with $|\Omega| \approx 10^{8}$.
The larger number of bounces quickly reveals that it is necessary to enforce the
Hamiltonian constraint. An explicitly constraint enforcing
$U(1)$ code was developed some years ago by Ove (see
\cite{ove90} and references therein).

It is well known \cite{belinskii73} that a scalar field can suppress Mixmaster
oscillations in homogeneous cosmologies. BKL argued that the suppression would
also occur in spatially inhomogeneous models. This was demonstrated numerically
for magnetic Gowdy and $U(1)$ symmetric spacetimes \cite{berger99a}. Andersson
and Rendall proved that completely general cosmological (spatially $T^3$)
spacetimes (no symmetries) with sufficiently strong scalar fields have generic
AVTD solutions \cite{andersson00}. Garfinkle \cite{garfinkle01b} has constructed
a 3D harmonic code which, so far, has found AVTD solutions with a scalar field
present. Work on generic vacuum models is in progress.

Cosmological models inspired by string theory contain higher derivative
curvature terms and exotic matter fields. Damour and Henneaux have applied the
BKL approach to such models and conclude that their approach to the singularity
exhibits LMD \cite{damour00a}.

Finally, there has been a study of the relationship between the ``long
wavelength approximation'' and the BKL analyses by Deruelle and Langlois
\cite{deruelle95}.

\section{Discussion}
Numerical investigation of singularities provides an arena for the close
coupling of analytic and numerical techniques. The references provided here
contain many examples of analytic results guided by numerical results and
numerical studies to demonstrate whether or not analytic conjectures are valid.

Even more striking is the convergence of the separate topics of this review.
While the search for naked singularities in the collapse of highly prolate
systems has yielded controversial results, a naked singularity was discovered in
the collapse of spherically symmetric scalar fields. The numerical exploration of
cosmological singularities has yielded strong evidence that the asymptotic
behavior is local---each spatial point evolves toward the singularity as a
separate universe. This means that conclusions from these studies should be
relevant in any generic collapse. This area of research then should begin to
overlap with the studies of black hole interiors (see for example
\cite{burko97b}).

\section*{Acknowledgements}
I would like to thank David Garfinkle for useful discussions (although any
errors are mine). This update was supported in part by National Science
Foundation Grant PHY9800103 to Oakland University.

\bibliography{numsing0601}

\end{document}